\documentclass[aps,epsfig,floats,twocolumn,showpacs]{revtex4}
\usepackage{epsfig}

\begin{document}

\title{Quantum phase slips in superconducting Nb nanowire networks deposited on self-assembled Si templates}
\author{C. Cirillo$^1$, M. Trezza$^1$, F. Chiarella$^1$, A. Vecchione$^1$, V. P. Bondarenko$^{2}$, S. L. Prischepa$^{2}$, and C. Attanasio$^1$}

\affiliation{$^1$CNR-SPIN Salerno and Dipartimento di Fisica \lq\lq E.R. Caianiello\rq\rq, Universit\`{a} degli Studi di Salerno, Fisciano (Sa) I-84084, Italy\\
$^2$Belarusian State University of Informatics and Radioelectronics, P. Browka 6, Minsk 220013, Belarus}

\date{\today}

\begin{abstract}

Robust porous silicon substrates were employed for generating interconnected networks of superconducting ultrathin Nb nanowires. Scanning electron
microscopy analysis was performed to investigate the morphology of the samples, which constitute of polycrystalline single wires with grain size
of about 10 nm. The samples exhibit nonzero resistance over a broad temperature range below the critical temperature, fingerprint of phase
slippage processes. The transport data are satisfactory reproduced by models describing both thermal and quantum fluctuations of the
superconducting order parameter in thin homogeneous superconducting wires.

\end{abstract}


\maketitle

The research activity in the field of superconductivity at reduced dimension has been continuously growing \cite{AruPhysRep,BezTopRev}, due to the
major implications that these results can have for understanding fundamental phenomena \cite{Johansson,Lau} as well as for possible applications
in superconducting electronics \cite{Astafiev,Ku,Nazarov,BezrPRB,BezNNW}. The behavior of the so-called superconducting nanowires, namely
ultrathin samples with dimensions comparable with the superconducting coherence length, $\xi$, is dominated by both thermal activated phase slips
(TAPS) and quantum phase slips (QPS) processes \cite{Giordano}, causing the wires to remain resistive much below the superconducting transition
temperature. The most challenging aspect of the experimental study of nanowires is the difficulty of fabricating homogeneous samples, since it has
been widely demonstrated that sample inhomogeneity can be the source of broadened superconducting transitions \cite{Bollinger,Zgirsti,Patel}. From
the early studies realized on crystalline superconducting whiskers \cite{Tinkham}, advances in nanofabrication techniques have allowed the
realization of high quality single crystals nanowires \cite{Wang}. Even more sophisticated is the approach of using suspended carbon nanotubes or
DNA molecules as templates for the formation of superconducting nanowires \cite{BezryadinNature,BezryadinScience,BezryadinAM}. A radically
different approach to nanostructures fabrication based on self-assembled growth attracted much attention also in the superconducting nanowire
field \cite{Yi,Michotte,Tian,Kwok}. Most of these works rely on the deposition within the channels and cavities of porous membranes.
Self-assembled methods are versatile and reliable bottom-up techniques for generating low-cost patterns of nanostructures, assuring a highly
reproducible geometry on very large areas.

%\begin{figure*}[!ht]
%\centering \includegraphics[height=0.20\textheight]{pori_new}\caption{Top-view scanning electron microscopy micrographs of (a) a free nanoporous
%Si substrate ($a= 50$ nm, $d= 15$ nm) and Nb nanowire networks deposited on PS with nominal thickness of (b) 3.5 nm, (c) 7 nm, (d) 15 nm. In the
%upper (lower) panels the images acquired at lower (higher) magnifications are reported. The white scale bar is 100 nm.}\label{imm}
%\end{figure*}

\begin{figure*}[!ht]
\centering \epsfig{figure=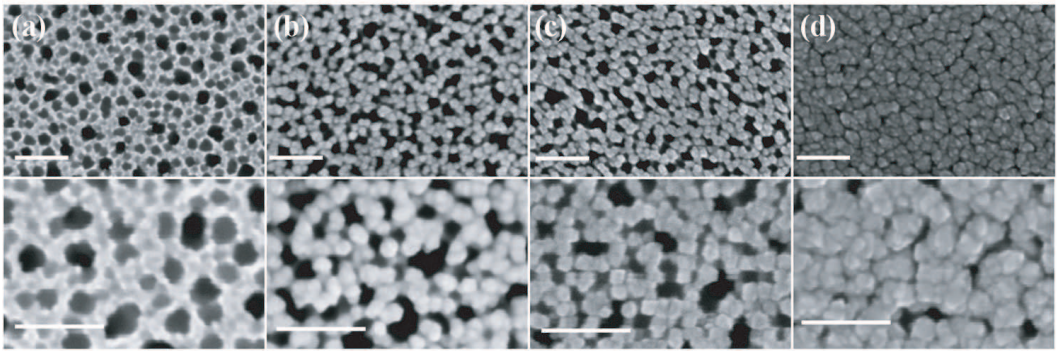,height=0.2\textheight,clip=}\caption{{\small \textit{Top-view scanning electron microscopy micrographs of
(a) a free nanoporous Si substrate ($a= 50$ nm, $d= 15$ nm) and Nb nanowire networks deposited on PS with nominal thickness of (b) 3.5 nm, (c) 7
nm, (d) 15 nm. In the upper (lower) panels the images acquired at lower (higher) magnifications are reported. The white scale bar is 100 nm.}}
\label{imm}}
\end{figure*}

In this work the formation of interconnected networks consisting of Nb ultrathin superconducting nanowires is achieved by using Porous Silicon
(PS) as template substrate. The extremely reduced film thickness favors the deposited material to occupy only the substrate pitch, therefore the
as-sputtered films result as a network of interconnected wires, whose average width, $w$, can be assumed to be equal to the periodic pore spacing
minus the pore diameter. Due to the extremely reduced characteristic dimension of the substrate $w$ is comparable to the superconducting coherence
length $\xi$, and hence each individual wire behaves as a one-dimensional (1D) object. As a consequence, the whole samples show broadened
resistive transitions, which can be described by theoretical models for both thermal and quantum fluctuations of the order parameter in 1D
superconductors \cite{Zaikin1,Zaikin2,Bae}. The proposed technique allows one to overcome the problem of handling fragile membranes usually used
at this purpose \cite{Yi,Michotte,Kwok} employing robust PS crystals as stable support. Moreover, the analyzed systems are rather simple and
macroscopically large objects, which, however, reveal fascinating quantum effects. Finally, compared to similar self-assembled nanowire networks
\cite{Kwok} exhibiting thermal phase slippage, the system presented in this work provides strong evidence of QPS. \\ \indent PS was fabricated by
electrochemical anodic etching of n-type, 0.01 $\Omega$cm, monocrystalline Silicon wafers as previously described in Refs. \cite{Trezza1,Trezza2}.
The resulting porous substrates, covering an area of about 2 cm$^2$, have average pore diameter $d= 10-20$ nm and average interpore spacing
(defined as the distance between the centers of two consecutive pores) $a= 20 - 50$ nm. Nb ultrathin films were deposited on top of PS substrates
in an UHV dc diode magnetron sputtering system with a base pressure in the low 10$^{-8}$ mbar regime following the same fabrication procedure
described elsewhere \cite{Trezza1,Trezza2}. Since the effect of the periodic template is reduced when the film thickness, $d_{Nb}$, exceeds the
pore diameter \cite{Trezza1}, the thickest Nb films deposited for transport measurements have $d_{Nb}= 12$ nm. With this approach an array of
interconnected Nb wires is guided by the PS template, the single wire width being the substrate pitch ($w = 10 - 35$ nm). Finally, the samples
were patterned by standard optical lithography and lift-off procedure into bridges of width $W_{b}=10, 20$ $\mu$m and length $L_{b}=100$ $\mu$m,
to obtain a classical pseudo-4-point geometry (meaning two contacts, each used for a current and a voltage lead). The effect of the patterning is
to reduce the number $N$ of interconnected wires under study.
\\ \indent Fig. \ref{imm} shows high resolution images of the surface of superconductive nanowire networks obtained from a template with $a= 50$
nm and $d= 15$ nm, performed by Field Emission Scanning Electron Microscopy (FESEM) ($\Sigma$igma Gemini by Zeiss). The images of four different
samples are reported in the panels: (a) S0, a free nanoporous Si substate, (b, c, d) three Nb films (S3.5, S7, and S15) with nominal thicknesses
$d_{Nb}= 3.5, 7$, and 15 nm, respectively. The images were acquired at two different magnifications (600 Kx and 1312 Kx) and clearly show the
evolution of the morphology with the progressive Nb deposition. It emerges that the single nanowires are polycrystalline with well shaped grains
with increasing dimension of 10 - 20 nm for increasing $d_{Nb}$ values. Images have been analyzed by SPIP \cite{SPIP}, a software for image
analysis, in order to evaluate the average diameter of pores, $d$, and grain size, $\zeta$. These value are summarized in Table \ref{table2}.

\begin{figure}[!ht]
\centering \epsfig{figure=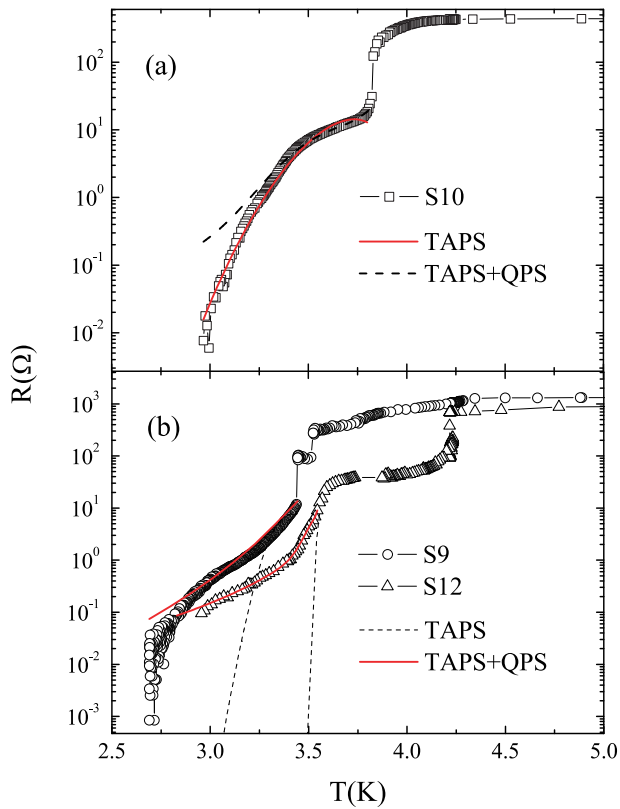,height=0.45\textheight,clip=}\caption{{\small \textit{(Color online). Resistive transitions, $R(T)$, of different
Nb nanowire networks: (a) squares represent the experimental data referring to sample S10, the solid red line is the theoretical curve obtained
including only the TAPS contribution in Eq. \ref{eq:Rtot}, while dashed black line is the result of the fitting procedure including both TAPS and
QPS contributions; (b) experimental data of samples S9 (circles) and S12 (triangles) are shown together with the curves obtained from Eq.
\ref{eq:Rtot} including only the TAPS term (dashed black lines) and both TAPS and QPS terms (solid red lines).}} \label{RT}}
\end{figure}

From the FESEM analyses, it follows that the optimal Nb thickness to investigate the superconducting properties of the nanowire array should lie
in the range $d_{Nb} \approx 9 - 12$ nm. This, in fact, assures that the wires are continuous, while preserving the presence of a well defined
network. Moreover, at these reduced thicknesses the system could approach the 1D limit. The characteristic parameters of these samples are listed
in Table \ref{table}.
\begin{table}
\begin{tabular}{cccc}
sample &$d_{Nb}$ (nm) & $\zeta$ (nm) & $d$ (nm)\\
\hline
S0 & 0 & - & 15 \\

S3.5 &3.5 & 10 & 13 \\

S7 &7 & 11 & 11 \\

S15 &15 & 20 & 6 \\
\end{tabular}
\vspace{4mm} \caption{Summary of the calculated morphological parameters at different film thickness ($d_{Nb}$). $\zeta$ indicates the grain size,
while $d$ is the pore diameter. Grains and pores average diameters are reported with an error of $\pm$2 nm.} \label{table2}
\end{table}
Superconducting transition temperatures, $T_{c}$, and critical currents, $I_{c}$, were resistively measured in a $^{4}$He cryostat using a
standard dc four-probe technique. Transport measurements provide also a probe of the quality and the homogeneity of the nanowires. It is worth
underlining that on the same kind of unstructured samples an extensive electrical characterization was performed, which revealed that ultrathin Nb
films deposited on PS substrates exhibit well established superconducting properties \cite{Trezza1,Trezza2} comparable with the ones of
conventional plane films. It has been also demonstrated that the size and spacing of PS pores is well matched to the vortex lattice of Nb thin
films, giving rise to matching effects at high magnetic fields and low temperature \cite{Trezza1,Trezza2}. The standard lift-off processing on a
micrometer scale is not expected to induce drastic sample imperfections which could be responsible of a dramatic resistance versus temperature,
$R(T)$, broadening. Resistivity of the single wires is difficult to estimate accurately due to the particular nanowires arrangements. However, on
first conservative approximation, it is reasonable to assume for $\rho_{N}$ a value comparable with the one obtained for plane Nb ultrathin films
\cite{Lehtinen}, which for sample deposited in the same sputtering system under the same conditions varies between $\rho_{N}= 50$ and 35 $\mu
\Omega$cm for thickness in the analyzed $d_{Nb}$ range \cite{rho}. Knowing the material constant $\rho_{N}$$\textit{l}$ for Niobium \cite{Minhaj},
namely $\rho_{N}$$\textit{l}= 3.72$ $\times$10$^{-6}$ $\mu \Omega$cm$^{2}$, it results that the value of the low temperature mean free path is
$\textit{l}$ $\approx$ 1 nm in agreement with the values reported for ultrathin Nb nanowires \cite{BezAPL}. Moreover, being $\textit{l}$ $\ll$
$\xi$, it follows that the samples are in the dirty limit, an important requirement for the occurrence of QPS processes, since the probability of
a QPS event can be expressed as $P \propto exp(-a T_{c}^{1/2} \sigma / \rho_{N})$ (here $\sigma \approx d_{Nb} \cdot w$ is the wire cross
sectional area) indicating that low temperature superconductors with high resistivity are the best candidates for observing QPS. Compared to
MoGe-based nanowire network \cite{Kwok} the detection of QPS is favored for Nb wires, since the required condition \cite{AruPhysRep} $k_{B} T \leq
\Delta(T)$, where $\Delta(T)$ is the temperature dependent superconducting gap, is realized at higher temperatures. Fig. \ref{RT} shows the
resistive transitions for all the analyzed samples. Measurements were performed using a constant bias current $I_{b}= 50$ $\mu$A. The first step
in the curves is due to the transition of the electrodes \cite{AruPhysRep,BezryadinNature}, therefore the values of $T_{c}$, reported in Table
\ref{table}, were evaluated at the midpoint of the transition occurring below the first one. Correspondingly, the normal state resistance,
$R_{N}$, is defined as the sample resistance below the electrodes transition \cite{AruPhysRep,BezryadinNature}. The main feature of the $R(T)$
curves is the nonzero resistance over a wide temperature range, which is strongly reminding of a 1D behavior. It also deserves noticing that the
$R(T)$ curves do not show any steps or humps, which can be a signature of inhomogeneity \cite{Bollinger}. Moreover, the moderate values of the
normal state resistance ($R_{N} \ll R_{Q}$, where $R_{Q}$ is the quantum resistance $R_{Q}=h/4e^{2} \approx 6.45$ k$\Omega$) exclude that weak
links constrictions or the film granularity can be responsible of the conduction \cite{Bollinger}. In order to further demonstrate the good
quality of the samples and to exclude the presence of tunneling barriers at the grain boundaries, it is useful to analyze both the shape of the
voltage-current ($V(I)$) characteristics and the values of their critical current density, $J_{c}$ \cite{BezryadinNature}. Indeed at low
temperature the shape of the $V(I)$ characteristics (not shown here) are very sharp and present no steps or hysteresis. The values of $J_{c}$ are
comparable with the ones measured on Nb perforated thin films \cite{JAPCara}. The critical current of the network is, indeed, $I_{c}=(N+1)i_{c}$,
where $i_{c}$ is the critical current of the single wire, which is related to its depairing current, $i_{dp}$, via
$i_{c}=2i_{dp}/3\sqrt{3}$.\cite{Zant} It follows that from the measured $I_{c}$ values, the depairing current density of the single nanowire at
$T=0$ can be estimated to be $J_{dp}= 1.5$ $\times$ 10$^{10}$ A/m$^{2}$, about one order of magnitude smaller than the depairing current density
evaluated in the framework of the model by Kupriyanov and Lukichev \cite{KL} (KL) $J^{KL}_{dp}= 3.5$ $\times$ 10$^{11}$ A/m$^{2}$, a discrepancy
comparable to the one reported in the case of Nb perforated thin films \cite{JAPCara}. Moreover, in the present case, this difference may be
attributed, for instance, to the approximation performed calculating the values of $N$, namely to the complex morphology of the samples.
\begin{table}
\begin{tabular}{cccccccccc}
sample  & $a$(nm) & $w$(nm) & $d_{Nb}$(nm) & $\sqrt{\sigma}$(nm) & $N$ & $W_{b}$($\mu$m) & $T_{c}$(K)\\
\hline
S9 & 20 & 10 & 9 & 9.5 & 500 & 10 & 3.39 \\

S10 & 40 & 30 & 10 & 17.3 & 250 & 10 & 3.52 \\

S12 & 40 & 30 & 12 & 19.0 & 500 & 20 & 3.57 \\
\end{tabular}
\vspace{4mm} \caption{Samples characteristic dimensions: $a$ indicates the interpore spacing, $w$ the wire width, $d_{Nb}$ the Nb thickness,
$\sqrt{\sigma}$ the wire effective diameter, $N$ the number of interconnected wires enclosed in the patterned bridge of width $W_{b}$, and $T_{c}$
the critical temperature of the samples. The numbers listed above are likely to be corrected considering a deviation from the corresponding mean
interpore distance of the order of 10\% \cite{Trezza1}. The pore diameter, $d$, is 10 nm for all the measured samples.} \label{table}
\end{table}
Having clarified the issue of the sample homogeneity, it is possible to analyze the $R(T)$ curves in the framework of the theoretical models
proposed for 1D superconductors to describe both thermal and quantum phase slip processes. Indeed the systems under study consist of
interconnected networks of ultrathin Nb nanowires with width ($w= 10, 30$ nm) comparable to the Nb superconducting coherence length. The latter
was estimated from the temperature dependence of the perpendicular upper critical field, $H_{c2\perp}(T)$, obtained performing resistance versus
field, $R(H)$, measurements at fixed values of the temperature. Indeed it follows that the Ginzburg-Landau coherence length at zero temperature is
$\xi(0)= 10$ nm. In presence of thermal activated phase slip processes the resistance is described by the equation \cite{Zaikin2}:

\begin{equation}
R_{TAPS}(T) = \frac{16\sqrt{6}}{\pi^{2/3}}R_{Q}\frac{L}{\xi(T)}\frac{T^{*}}{T}\sqrt{\frac{U(T)}{k_{B}T}}\exp\Bigg[-\frac{U(T)}{k_{B}T}\Bigg]
\label{eq:TAPS}
\end{equation}

\noindent while QPS contribution may be expressed with exponential accuracy as \cite{Zaikin1,Bae,Lehtinen}:

\begin{equation}
R_{QPS}(T) \approx AB\frac{R_{Q}^{2}}{R_{N}}\frac{L^{2}}{\xi^{2}(0)}\exp\Bigg[-A\frac{R_{Q}}{R_{N}}\frac{L}{\xi(T)}\Bigg] \label{eq:QPS}
\end{equation}

\noindent where $T^{*} \approx T_{c}$ is a crossover temperature between TAPS and QPS regime, $L$ is the nanowire length, and
$\xi(T)=\xi_{0}/\sqrt{1-T/T_{c}}$ and $U(T)=U(0)(T_{c}-T)^{3/2}$ are the temperature dependent coherence length and phase slip activation energy,
respectively. Since $L$ is not well defined in these samples \cite{Kwok}, it is treated as fitting parameter together with $T_{c}$ and, in the
case of Eq. \ref{eq:TAPS} (\ref{eq:QPS}), the parameter(s) $U(0)$ ($A$ and $B$), in the framework of the same approach used in Refs.
\cite{Kwok,Bae}. Thus, the total resistance of the samples can be expressed as \cite{Lau}:

\begin{equation}
R(T) = [R_{N}^{-1} + (R_{TAPS}+R_{QPS})^{-1}]^{-1}  \label{eq:Rtot}
\end{equation}

The results of the theoretical analysis are reported in Fig. \ref{RT}. For the sake of the clarity the best fitting curves are indicated by solid
red lines. The data corresponding to sample S10 [Fig. \ref{RT}(a)] present a negative curvature consistent with thermal activation process, namely
the $lnR(T)$ curve does not level off at lower $T$ as characteristic of quantum tunneling and they were satisfactorily fitted retaining only the
TAPS term (Eq. \ref{eq:TAPS}) in Eq. \ref{eq:Rtot}. Indeed, sample S10 is characterized by a slightly lower $R_{N}$ value and a slightly higher
$T_{c}$ value compared to S9 and S12. In fact, from Fig. \ref{RT}(a) it can be inferred that including also the QPS term (Eq. \ref{eq:QPS}) in Eq.
\ref{eq:Rtot} produces a much broader transition (dashed black line).

%\begin{figure}[!ht]
%\centering \includegraphics[height=0.35\textheight]{RT}\caption{(Color online). Resistive transitions, $R(T)$, of different Nb nanowire networks:
%(a) squares represent the experimental data referring to sample S10, the solid red line is the theoretical curve obtained including only the TAPS
%contribution in Eq. \ref{eq:Rtot}, while dashed black line is the result of the fitting procedure including both TAPS and QPS contributions; (b)
%experimental data of samples S9 (circles) and S12 (triangles) are shown together with the curves obtained from Eq. \ref{eq:Rtot} including only
%the TAPS term (dashed black lines) and both TAPS and QPS terms (solid red lines).} \label{RT}
%\end{figure}

The results of the theoretical interpretation according to Eq. \ref{eq:Rtot} for samples S9 and S12 are reported in Fig. \ref{RT}(b) by thick red
lines. As far as the sample S12 is concerned, the best fit to the $R(T)$ curve obtained with Eq. \ref{eq:Rtot} nicely follows the data, which, on
the other hand strongly deviate from the solely TAPS dependence obtained from the same equation disregarding the QPS term (dashed black line).
These results suggest that the pronounced resistance tail exhibited by sample S12 could indeed originate from QPS processes. On the other hand the
$R(T)$ dependence of bridge S9 arising from both TAPS and QPS (Eq. \ref{eq:Rtot}) is in agreement with the experimental points only down to $T
\approx 2.75$ K. Below this temperature the curvature of the $lnR(T)$ curve changes again with a slope more consistent with thermal activation.
However, the discrepancy between the TAPS dependence obtained from Eq. \ref{eq:Rtot} considering only the thermal contribution (dashed black line)
and the experimental data is extremely pronounced. It is worth commenting the values of the parameters extracted from the fitting procedure. In
particular the values of the critical temperature, namely $T_{c} = 3.5, 4.0, 3.5$ K, estimated for samples S9, S10, and S12, respectively, are
close to the measured ones (see Table \ref{table}). The best fitting procedures performed with the full Eq. \ref{eq:Rtot} was obtained using $B
\approx 2 - 40$, in agreement with the values reported in the literature \cite{Bae}, but at the price of considering quite high values of $A
\approx 1000$. This discrepancy can be possibly attributed to the exponential approximation assumed in treating the QPS contribution. Finally for
sample S10 a value of $U(0)=2.5$ meV was extracted \cite{Kwok}. From the relation $U(0) \approx 0.83LR_{Q}T_{c}/\xi(0)R_{N}$ valid for a single
nanowire in the framework of the approximate phenomenological formula of Arrhenius-Little \cite{Bae}, following Ref. \cite{Kwok}, it is possible
to estimate $L \approx 312$ nm. \\ \indent This analysis reveals that, despite the models were derived for an individual wire, they can reproduce
the transition of networks with not well defined length and, moreover, with finite widths and activation energies distribution. Clearly these
spreads are not taken into account in these models, as well as the distribution of the strengths of the contacts between the nanowires in the
network. Indeed poor contacts, which may be present in these systems for instance at grain boundaries, can behave such weak links, the presence of
which increases the probability of TAPS and QPS \cite{Bae}. These parameters fluctuations, which are due to the peculiar growth technique of the
network, are expected to affect more the samples with lower values of $\sigma^{1/2}$. In this sense the $R(T)$ behavior of sample S12, which is
expected to be more homogeneous compared to sample S9, is better reproduced by Eq. \ref{eq:Rtot}. In this respect, however, it is worth commenting
that, since in arrays of small Josephson contacts the thermal activation of the Josephson weak links may also affect the shape of the $R(T)$
curves, the experimental data were also analyzed in the framework of the theory of Ambegaokar and Halperin \cite{AmbHal} within the formulation
given in Ref. \cite{BezrPRB}. The procedure revealed an extremely poor agreement between the $R(T)$ dependencies and the model for any reasonable
physical value of the fitting parameters, confirming that weak links do not dominate the transport in these systems. Finally, to shed a light on
the physics revealed by these systems two complementary experiments could be performed. On one hand electron beam lithography could be used to
fabricate much narrower bridges, obtaining better controlled arrays with a drastically reduced number of interconnected wires and, consequently, a
narrower distribution of widths and contact strength. Moreover the analysis of the microwave response of the arrays could help to determine the
dominant dissipation process, namely TAPS or QPS \cite{Bae}.

In summary, both thermal and quantum fluctuations were revealed from $R(T)$ measurements in superconducting Nb nanowire networks patterned on PS
substrates. The innovation of the analyzed systems lies both in the widely accessible fabrication technique, which rely on rigid templates, and,
most importantly, in the multiple-connectivity of the wires. \\ \indent The research leading to these results has received funding from the
European Union Seventh Framework Programme (FP7/2007- 2013) under grant agreement N. 264098 – MAMA. The authors wish to thank Prof. A. D. Zaikin
for useful suggestions.


\vspace{0.5in}



\begin{references}

\bibitem{AruPhysRep} K.Y. Arutyunov, D.S. Golubev, and A.D. Zaikin, Physics Reports {\bf 464}, 1 (2008).

\bibitem{BezTopRev} A. Bezryadin, J. Phys.: Condens. Matt. {\bf 20}, 043202 (2008).

\bibitem{Johansson} A. Johansson, G. Sambandamurthy, D. Shahar, N. Jacobson, and R. Tenne, Phys. Rev. Lett. {\bf95}, 116805 (2005).

\bibitem{Lau} C.N. Lau, N. Markovic, M. Bockrath, A. Bezryadin, and M. Tinkham, Phys. Rev. Lett. {\bf87}, 217003 (2001).

\bibitem{Astafiev} O.V. Astafiev, L.B. Ioffe, S. Kafanov, Yu.A. Pashkin, K.Yu. Arutyunov, D. Shahar, O. Cohen, and J.S. Tsai, Nature {\bf484}, 355 (2012).

\bibitem{Ku} J. Ku, V. Manucharyan, and A. Bezryadin, Phys. Rev. B {\bf82}, 134518 (2010).

\bibitem{Nazarov} J.E. Mooij and Yu.V. Nazarov, Nat. Phys. {\bf2}, 169 (2006).

\bibitem{BezrPRB} D. Pekker, A. Bezryadin, D.S. Hopkins, and P.M. Goldbert, Phys. Rev. B {\bf308}, 1762 (2005).

\bibitem{BezNNW} A. Bezryadin, Nature {\bf484}, 324 (2012).

\bibitem{Giordano} N. Giordano, Phys. Rev. B {\bf41}, 6350 (1990).

\bibitem{Bollinger} A.T. Bollinger, A. Rogachev, M. Remeika, and A. Bezryadin, Phys. Rev. B {\bf69}, 180503(R) (2004).

\bibitem{Zgirsti} M. Zgirsti and K.Yu. Arutyunov, Phys. Rev. B {\bf75}, 172509 (2007).

\bibitem{Patel} U. Patel, Z. L. Xiao, A. Gurevich, S. Avci, J. Hua, R. Divan, U. Welp, and W. K. Kwok, Phys. Rev. B {\bf80}, 012504 (2009).

\bibitem{Tinkham} R.S. Newbower, M.R. Beasley, and M. Tinkham, Phys. Rev. B {\bf5}, 864 (1972).

\bibitem{Wang} J. Wang, X.-C. Ma, L. Lu, A.-Z. Jin, C.-Z. Gu, X.C. Xie, J.-F. Jia, X. Chen, and Q.-K. Xue, Appl. Phys. Lett. {\bf92}, 233119 (2008).

\bibitem{BezryadinNature} A. Bezryadin, C.N. Lau, and M. Tinkham, Nature {\bf404}, 971 (2000).

\bibitem{BezryadinScience} D.S. Hopkins, D. Pekker, P.M. Goldbert, and A. Bezryadin, Science {\bf308}, 1762 (2005).

\bibitem{BezryadinAM} A. Bezryadin and P.M. Goldbart, Adv. Mater. {\bf22}, 1111 (2010).

\bibitem{Yi} G. Yi and W. Schwarzacher, Appl. Phys. Lett. {\bf74}, 1746 (1999).

\bibitem{Michotte} S. Michotte, L. Piraux, S. Dubois, F. Pailloux, G. Stenuit, and J. Govaerts, Physica C {\bf377}, 267 (2002).

\bibitem{Tian} M. Tian, J. Wang, J.S. Kurtz, Y. Liu, M.H.W. Chan, T.S. Mayer, and T.E. Mallouk, Phys. Rev. B {\bf71}, 104521 (2005).

\bibitem{Kwok} Q. Luo, X.Q. Zeng, M.E. Miszczak, Z.L. Xiao, J. Pearson, T. Xu, and W.K. Kwok, Phys. Rev. B {\bf85}, 174513 (2012).

\bibitem{Zaikin1} A.D. Zaikin, D.S. Golubev, A. van Otterlo, and G.T. Zim$\acute{\mathrm{a}}$nyi, Phys. Rev. Lett. {\bf78}, 1552 (1997); D.S. Golubev and A.D. Zaikin, Phys. Rev. B {\bf64}, 014504 (2001).

\bibitem{Zaikin2} D.S. Golubev and A.D. Zaikin, Phys. Rev. B. {\bf78}, 144502 (2008).

\bibitem{Bae} M.-H. Bae, R.C. Dinsmore III, T. Aref, M. Brenner, and A. Bezryadin, Nano Lett. {\bf9}, 1889 (2009).

\bibitem{Trezza1} M. Trezza, S.L. Prischepa, C. Cirillo, R. Fittipaldi, M. Sarno, D. Sannino, P. Ciambelli, M.B.S. Hesselberth, S.K. Lazarouk, A.V. Dolbik, V.E.
Borisenko, and C. Attanasio, J. Appl. Phys. {\bf104}, 083917 (2008).

\bibitem{Trezza2} M. Trezza, C. Cirillo, S.L. Prischepa, and C. Attanasio, Europhys. Lett. {\bf88}, 57006 (2009).

\bibitem{SPIP} G. Ausanio, V. Iannotti, S. Amoruso, X. Wang, C. Aruta, M. Arzeo, R. Fittipaldi, A. Vecchione, R. Bruzzese, and L. Lanotte, Appl. Surf. Sci. {\bf258}, 9337 (2012).

\bibitem{Lehtinen} J.S. Lehtinen, T. Sajavaara, K.Yu. Arutyunov, M.Yu. Presnjakov, and A.L. Vasiliev, Phys. Rev. B. {\bf85}, 094508 (2012).

\bibitem{rho} C. Cirillo and C. Attanasio, (private communication).

\bibitem{Minhaj} M.S.M. Minhaj, S. Meepagala, J.T. Chen, and L.E. Wenger, Phys. Rev. B  {\bf49}, 15235 (1994).

\bibitem{BezAPL} A. Rogachev and A. Bezryadin, Appl. Phys. Lett. {\bf83}, 512 (2003).

\bibitem{JAPCara} P. Sabatino, C. Cirillo, G. Carapella, M. Trezza, and C. Attanasio, J. Appl. Phys. {\bf108},  053906 (2010).

\bibitem{Zant} H.S.J. van der Zant, M.N. Webster, J. Romijn, and J.E. Mooij, Phys. Rev. B. {\bf50}, 340 (1994).

\bibitem{KL} M.Y. Kupriyanov and V.F. Lukichev, Fiz. Nizk. Temp. {\bf6}, 445 (1980) [Sov. J. Low Temp. Phys. {\bf6}, 210 (1980)].

\bibitem{AmbHal} V. Ambegaokar and B.I. Halperin, Phys. Rev. Lett. {\bf22}, 1364 (1969).

\end{references}
\end{document}